\documentclass[traditabstract]{aa}

\usepackage{amsfonts}
\usepackage{amsmath}
\usepackage{amssymb}
\usepackage{pdflscape}
\usepackage{rotating}
\usepackage{natbib}
\usepackage{adjustbox}
\usepackage{txfonts}
\usepackage{longtable}
\usepackage[small,bf]{caption}
\usepackage[small,bf]{subfigure}
\usepackage{boldline,multirow}
\usepackage{xspace}
\usepackage{graphicx}
\usepackage{epstopdf}
\usepackage[flushleft]{threeparttable}
\usepackage[pdftitle={3c120}, colorlinks=true, citecolor=blue, linkcolor=blue, urlcolor=blue]{hyperref}

\newcommand\simlt{\lower.5ex\hbox{$\; \buildrel < \over \sim \;$}}
\newcommand{\fermi}{{\it Fermi}-LAT\xspace}
\newcommand{\gray}{$\gamma$-ray\xspace}
\newcommand{\grays}{$\gamma$-rays\xspace}

\DeclareUnicodeCharacter{B0}{\textdegree}

\begin{document}
\title{ Fermi-LAT Observation of Non-Blazar AGNs}
\titlerunning{\grays from non-blazar AGNs}

\author{N. Sahakyan\inst{1,2}
\and  V. Baghmanyan\inst{2,3}
\and D. Zargaryan\inst{2}
}
 \institute{ICRANet, Piazza della Repubblica 10, I-65122 Pescara, Italy
 \and ICRANet Armenia, Marshall Baghramian Avenue 24a, Yerevan 0019, Republic of Armenia
 \and Yerevan State University, Alek Manukyan Str. 1, Yerevan, 0025, Armenia }

  \abstract
   {{\it Fermi}-LAT has recently detected $\gamma$-ray emission from active galactic nuclei which do not show clear evidence for optical blazar characteristics or have jets pointing away from the observer (non-blazar AGNs). These are interesting $\gamma$-ray emitters providing an alternative approach to studying high energy emission processes.}
   { Investigation of the spectral and temporal properties of $\gamma$-ray emission from non-blazar AGNs using the recent {\it Fermi}-LAT observational data.}
   {The data collected by {\it Fermi}-LAT during 2008-2015, from the observations of 26 non-blazar AGNs, including eleven FRI and ten FRII radio galaxies/SSRQs and five NLSY1s are analyzed using the new PASS 8 event selection and instrument response function. Possible spectral changes above GeV energies are investigated with a detailed spectral analysis. Light curves generated with normal and adaptive time bins are used to study the $\gamma$-ray flux variability.}
   {Non-blazar AGNs have a \gray photon index in the range of $1.84-2.86$ and a flux varying from a few times $10^{-9}\:{\rm photon\:cm^{-2}s^{-1}}$ to $10^{-7}\:{\rm photon\:cm^{-2}s^{-1}}$. Over long time periods, power-law provides an adequate description of the \gray spectra of almost all sources. Significant curvature is observed in the \gray spectra of NGC 1275, NGC 6251, SBS 0846+513 and PMN J0948+0022 and their spectra are better described by log-parabola or power-law with exponential cut-off models. The \gray spectra of PKS 0625-25 and 3C 380 show a possible deviation from a simple power-law shape, indicating a spectral cutoff around the observed photon energy of $E_{\rm cut}=131.2\pm88.04$ GeV and $E_{\rm cut}=55.57\pm50.74$ GeV, respectively. Our analysis confirms the previous finding of an unusual spectral turnover in the $\gamma$-ray spectrum of Cen A: the photon index changes from $\Gamma=2.75\pm0.02$ to $2.31\pm0.1$ at $2.35\pm0.08$ GeV.  In the $\Gamma-L_{\gamma}$ plane, the luminosity of non-blazar AGNs is spread in the range of $(10^{41}-10^{47})\:{\rm erg\:s^{-1}}$, where the lowest luminosity have FRI radio galaxies (but typically appear with a harder photon index) and the highest- SSRQs/NLSY1s (with softer photon indexes). We confirm the previously reported short-timescale flux variability of NGC 1275 and 3C 120. The $\gamma$-ray emission from NLSY1s, 1H 0323+342, SBS 0846+513 and PMN J0948+0022, is variable, showing flares in short scales sometimes accompanied by a moderate hardening of their spectra (e.g., on MJD 56146.8 the $\gamma$-ray photon index of SBS 0846+513 was $\Gamma=1.73\pm0.14$). 3C 111,  Cen A core, 3C 207, 3C 275.1, 3C 380, 4C+39.23B, PKS 1502+036 and PKS 2004-447 show a long-timescale flux variability in the $\gamma$-ray band.}
{}
 \maketitle
\section{Introduction}\label{sec1}
The recent observations in the High Energy (HE; $100\:{\rm MeV}<E_{\gamma}<100\:{\rm GeV}$) \gray band show that the extragalactic \gray sky is dominated by the emission from Active Galactic Nuclei (AGN) of different types. The majority of detected sources are blazars (BL Lacertae objects (BL-Lacs) and Flat-Spectrum Radio Quasars [FSRQs])- an extreme class of AGNs with jets forming a small angle with respect to the line of sight,  which makes their emission strongly Doppler boosted \citep{urry}. Blazars are known to emit electromagnetic radiation in almost all frequencies that are currently being observed, extending from radio to very high energy (VHE; >100 GeV) \gray bands. Their broadband spectra are mainly dominated by the non-thermal emission from a relativistic jet pointing toward the observer. This non-thermal emission is characterized by variability in all observed energy bands with different variability time scales ranging from years down to a few minutes. The shortest variability time scales are usually observed in the HE and VHE \gray bands, e.g., the rapid changes of the \gray flux of PKS 2155-304, IC 310, PKS 1510-089, 3C 454.3, 3C 273, PKS B1222 + 216, etc. \citep{aharpks,alek310, brown13,Foschini11,Foschini13, Saito13, nalewajko, hnm15, Rani13, aaa16}. This implies that the emission is produced in a very compact region which provides a unique chance to investigate the jet structure and emission from sub-parsec regions.\\
Besides blazars, the $\gamma$-ray emission from other types of AGNs, e.g., which do not show a clear evidence for optical blazar characteristics or have jets pointing away from the observer (non-blazar AGNs) also have been detected. These sources were not considered as favored GeV emitters because the nonthermal emission from their jets is less Doppler boosted. After the launch of Fermi Large Area Telescope (\fermi), several non-blazar AGNs (mostly radio galaxies) were detected already in the first year of observations \citep{abdocena,abdom87,abdongc1275, 1FGL}. This provided a chance to investigate also the emission from extended non-boosted regions, moderately relativistic plasmas, etc.\\
Analyzing the \fermi data accumulated during the first fifteen months of observations of non-blazar AGNs (including seven Fanaroff-Riley Type I (FRI) radio galaxies \citep{fanaroff74} and four FRII radio sources consisting of two FRII radio galaxies and two steep spectrum radio quasars [SSRQs]), \citet{abdomisaligned} showed that their average \gray spectral indexes are comparable, but they are somewhat less \gray luminous than their parent population of BL Lacs and FSRQs. Despite the fact that a few non-blazar AGNs had been detected, which prevented extensive population studies, the observations of individual sources provided new and interesting results. Most interesting is perhaps the detection of \gray emission from the lobes of Centaurus A (Cen A) and Fornax A \citep{abdolobes, fermiFornax} radio galaxies. In the case of Cen A, the emission extends up to $\approx3$ GeV with a soft photon index of $\Gamma\simeq 2.6$ and its flux makes more than half of the total source emission \citep{abdolobes,yang}. Instead, the \gray emission from the lobes of Fornax A has a power-law photon index of $\Gamma\simeq2.0$ and extends up to $20$ GeV. It illustrates that the bipolar or quasi-symmetrical structures in the radio maps around almost all the radio galaxies can be potential  sources of \grays.\\ 
The \gray spectra of some radio galaxies appeared with interesting features above several GeV. For example, during the \gray flaring activity of NGC 1275 in April-May 2009, a hysteresis behavior in the flux versus photon index plane was observed \citep{kataokangc1275}, when the \gray photon index hardened from $\Gamma\simeq2.2$ to $\Gamma\simeq2.0$. Next, during its major \gray flares of October 2015 and December 2016/January 2017, in the flux versus photon index plane the spectral evolution followed correspondingly a counter clockwise and a clockwise loop inferred from the light curve generated by an adaptive binning method \citep{baghmanyan17}. Another example is the second component in the \gray spectrum of the Cen A core emission; at $E_b \simeq (4.00\pm0.09)$ GeV the photon index changes from $\Gamma_1=2.74\pm0.03$ to $\Gamma_2=2.09\pm0.20$ \citep{sahakyancen}.\\
The multiwavelength observations of two well-known radio galaxies, 3C 111 and 3C 120, which show superluminal apparent motions in their radio jets, provided some clue on the localization of the emitting region. From 2008 to 2010, 3C 111 had only occasionally been detected in the HE \gray band. It was bright at the end of 2008, and faint for the rest of the time \citep{grandi}. These bright periods in the \gray band coincide with the millimeter-optical-X-ray outburst, suggesting cospatiality of the events, which allowed to localize the \gray dissipative zone to be at a distance of about $\sim 0.3$ pc from the black hole in a compact region confined within $\sim 0.1$ pc \citep{grandi}. Similarly, very long baseline interferometry and very long baseline array monitoring of 3C 120 reveal that its \gray activity is accompanied by the radio and millimeter core brightening followed by the ejection of a superluminal knot \citep{casadio,Tanaka3c120}. In \citet{casadio} it has been shown that \grays are detected only when the components are moving in a direction closer to our line of sight. The analysis of \gray events detection and superluminal components ejection periods shows that the \gray emission region is within $\sim0.13$ pc from the millimeter VLBI core \cite{casadio}.\\
The origin of the \gray emission from non-blazar AGNs is another puzzling problem. In the case of blazars the fast variability indicates the emission is produced in a compact region of the jet which narrows the range of possible emission mechanisms. For some non-blazar AGNs, their jets are less Doppler-boosted, thus the \gray emission from other potential sites (e.g., moderately relativistic outflows, radio lobes, accretion regions, etc.) can be detected as well, making exact theoretical modeling of the origin of emission harder. Even when the emission is most likely produced in a compact region of the jet (e.g., Cen A core \citep{abdocena}, M87 \citep{abdom87} and NGC 1275 \citep{abdongc1275}), the most popular one-zone synchrotron/synchrotron self Compton (SSC) model clearly fails to explain the observed multiwavelength properties.  More complicated models, multi-zone \citep{lenain} or structured jet \citep{tavecchio87,janiak,tavecchio14}, based on different assumptions should be employed to explain the observed data properly. New and high-quality data are still needed to fully understand the origin of multiwavelength emission from non-blazar AGNs.\\ 
We note that the emission from some non-blazar AGNs extends up to the VHE \gray band, allowing to detect them by current-generation of ground- based detectors. The observations of non-blazar AGNs in MeV/GeV (HE) and > GeV (VHE) bands, allowed to examine the nonthermal emission processes in the jets of AGNs which are not strongly affected by Doppler boosting. Up to now only the spectra of Cen A, M87, NGC 1275 and PKS 0625-35 have been measured in the MeV/GeV and TeV bands. The recent simulations show that at least eight new non-blazar AGNs will be detected by CTA at $\sim10\:\sigma$ under the assumption of a straight extrapolation of the \fermi measured power-law spectra in the CTA energy range \citep{angioni}. In this view, it is important to investigate the non-blazar AGNs \gray spectra above $\sim$GeV, which now can be done with much improved statistics than before, considering {\it i)} the \fermi data from longer periods of observation are collected and available and {\it ii)} the recent update of the event-level analysis from {\it PASS7} to {\it PASS8}. Moreover, using longer periods, it will be possible to study the \gray flux variability in both long and short time scales. This is crucial for theoretical modeling and distinguishing between compact and extended emission regions. This has motivated us to have a new look on the \gray emission from non-blazar AGNs.\\
In this paper we report on a detailed analysis of the \gray emission from 27 non-blazar AGNs which are included in the third catalog of AGNs detected by \fermi \citep{3fgl} (Section \ref{sec1.1}). In Section \ref{sec2}, the \fermi observations and the data reduction procedure are described. The results of the analysis are given in Section \ref{sec2.2}. The discussions and conclusions are presented in Sections \ref{sec3} and \ref{sec4}, respectively.
\section{Sources}\label{sec1.1}
The majority of AGNs detected by \fermi are blazars, FSRQs and BL Lacs, and the number of unknown-type blazars is almost the same as that of FSRQs ($\sim460$) \citep{3fgl}. Only a very small fraction of the total sample ($2\%$) are non-blazar AGNs, including fourteen radio galaxies, six SSRQs, one Compact Steep-Spectrum Quasar (CSSs), five Narrow Line Seyfert 1s (NLSy1s) and five other AGNs. From this list, the other AGNs show spectral energy distribution (SEDs) typical for radio-loud compact-core objects, but data are lacking to be more specific about their classes, thus we have excluded them from our sample. Other sources have well-established classification and/or have a well evaluated optical spectrum (with clear evidence for or lack of emission lines) and some of them are also related with blazars in the unification scenario proposed for radio-loud AGNs; e.g., FRIs are the parent population of BL Lacs, while FRIIs are linked with FSRQs \citep{urry}. We included all these objects in our source sample, i.e., AGNs with relativistic jets which are not blazars, and we refer to them collectively as non-blazar AGNs. Even if some of these sources might have  different jet properties (e.g., the jets of 3C 111 and 3C 120 with superluminal apparent motions, or the jets of some objects have a smaller but still larger viewing angle than that of blazars) we did not exclude them from our source sample to have a general view of the comparison of \gray emission characteristics of blazar and non-blazar objects.\\ 
The radio galaxies detected in the \gray band so far can be sub-divided into two morphological types, FRI and FRII \citep{fanaroff74}, according to their radio luminosities. FRIs have a typical luminosity of $< 10^{41} {\rm erg\:s^{-1}}$ (at 178 MHz) while that of the luminous FRIIs is $> 10^{41} {\rm erg\:s^{-1}}$ (at 178 MHz). FRIs have slower, more turbulent and less collimated jets, while FRIIs have powerful collimated jets (often only one is visible) which terminate into well-defined lobes with prominent hot spots. An example of an FRI is the nearby radio galaxy Cen A, and Pictor A with its peculiar jet is an example of FRII. Increasingly aligned versions of FRIIs are SSRQs, which are powerful radio sources with large-scale radio structures and which appear at intermediate angles between FRIIs and FSRQs. SSRQs are sometimes classified as CSS quasars, since most of the radio flux is emitted within galactic scales ($< 10$ kpc) rather than at hundreds of kilo-parsec scales \citep{gelderman}. Their radio structure on milliarcsec scales is similar to the morphology of a large radio source with lobes, hot spots and jets; because of this similarity and the observed high radio power, these sources are thought to be the precursors of large radio galaxies observed at the early stage of their expansion \citep{fanti,odea}. In the third catalog of AGNs detected by \fermi, only 4C +39.23B has been classified as a CSS quasar although SSRQs 3C 286 and 3C 380 also have characteristics similar to CSSs.\\
Recently, NLSy1s have also been classified as \gray emitters \citep{abdoNLSY}. NLSy1s are AGNs with optical spectral properties similar to those of Seyfert 1 galaxies, except for having narrow Balmer lines and strong optical ${\rm F_eII}$ lines \citep{ostrobock}. NLSy1 exhibit strong X-ray variability, steep X-ray spectra, relatively high luminosity, and a substantial soft X-ray excess \citep{Gallo}. These characteristics point to systems with a central black hole having smaller mass ($10^6 - 10^8\; M_{\odot}$) and higher accretion rates (close to or above the Eddington limit). Up to now it is not clear whether or not the jets of NLSy1s make a small angle to the observer's line of sight, however, they are very bright and interesting \gray emitters. Including them in our sample allows us to compare their \gray spectra, flux, luminosity, etc. with those of radio galaxies and blazars (aligned and not aligned cases) which facilitates their jet studies.\\
In our sample of non-blazar AGNs we did not include Fornax A and IC 310. In the first source the \grays are associated with its radio lobes \citep{fermiFornax}, while the second is most likely a BL Lac \citep{kadler}. In addition to the non-blazar AGNs from \citet{3fgl} we added also 3C 120, as several different groups have independently reported on the detection of constant \gray emission from it \citep{sahak3c120,casadio,Tanaka3c120}. Therefore there are in total 27 non-blazar AGNs considered in our study (see Table \ref{tab:a}).
\section{\fermi Data Analysis}\label{sec2}
The LAT on board the Fermi satellite is a pair-conversion \gray telescope that observes photons with energies between 20 MeV and 300 GeV. Launched on June 11 2008, it is in the default "all-sky-survey" mode optimized to provide a relatively uniform coverage of the entire sky every three hours. Further details of the instrument are described in \citet{atwood09} and references therein.
\subsection{Data extraction}
The data set used here was collected during the first 7 years of \fermi operation, from August 4 2008 to August 4 2015. We selected only events within a maximum zenith angle of $90^{\circ}$ to reduce the Earth limb \grays and use the recommended filter string "DATA\_ Qual$>$0 \&\& LAT\_ CONFIG==1" in {\it gtmktime}. The standard Fermi Science Tools v10r0p5 software package released on May 18 2015, available from the Fermi Science Support Center \footnote{http://fermi.gsfc.nasa.gov/ssc/data/analysis/software/} is used with the P8R2\_ SOURCE\_ V6 set of instrument response functions. Only events with higher probability of being photons (evclass=128 evtype=3) have been used within the energy range from 100 MeV to 300 GeV.\\
The photons from a circular region with a radius of $10^{\circ}$ around each source under consideration were used, except for 3C 207, M87, PKS 1502+036, PKS 2004-447 and 1 H 0323+342, around which  a region with a radius of $12^{\circ}$ was used for a better representation of the region of interest (RoI). By selecting different radii of the RoI we ensure that the selected RoIs are an accurate representation of the observation. The photons are binned with {\it gtbin} tool with a stereographic projection into pixels of  $0.1^{\circ}\times0.1^{\circ}$ and into 35 equal logarithmically spaced energy bins. The model for which we calculated the likelihood includes a combination of point-like and diffuse sources within a $14^{\circ}\times14^{\circ}$ square RoI ($16.9^{\circ}\times16.9^{\circ}$ when the radius of the circular region is $12^{\circ}$). The model file for each source is created using the \fermi third source catalog \citep{acero} (3FGL) where all sources falling between RoI and RoI+$5^{\circ}$ region are included in it. The Galactic and diffuse background models {\it gll\_ iem \_ v06} and {\it  iso\_P8R2\_SOURCE\_V6\_v06} have been used, which are currently recommended by the LAT team. The normalization of background models as well as fluxes and spectral indexes of the sources within RoI are left as free parameters in the analysis while for the sources outside RoI, the spectral slopes and normalizations were fixed to the values given in the 3FGL catalog.
\subsection{Spectral analysis}
 In order to find the best matches between spectral models and events, the binned likelihood analysis is performed with {\it gtlike} tool. The detection significance of each source is quantitatively estimated using the Test Statistics (TS) defined as TS = 2(log $L$ - log $L_0$), where $L$ and $L_0$ are the likelihoods with an additional source and null hypothesis \citep{mattox}. The spectra of the considered sources are modeled with a Power-Law (PL) model in the form of
 \begin{equation}
\frac{dN}{dE_{\gamma}}= N_{0}\:\left(\frac{E_{\gamma}}{E_{0}}\right)^{-\Gamma},
\label{PL}
\end{equation}
where the normalization $N_{0}$ and \gray photon index $\Gamma$ are considered as free parameters. If deviation from the PL model is found, then a more complex Power Law with Exponential Cut-off (PLEC) model in the form of
 \begin{equation}
\frac{dN}{dE_{\gamma}}= N_{0}\:\left(\frac{E_{\gamma}}{E_{0}}\right)^{-\Gamma} Exp\left(-\left(\frac{E_{\gamma}}{E_{c}}\right)^{\beta}\right),
\label{PLEC}
\end{equation}
is used, where $N_{0}$, $\Gamma$ and $E_{\rm c}$ are free parameters. First we fix the value of $\beta=1$, then consider it also as a free parameter. Since the Log-Parabolic (LP) spectral shape gave a better modeling for the spectra of some blazars \citep{massaro}, we consider also the LP model in the form of
\begin{equation}
\frac{dN}{dE_{\gamma}}= N_{0}\:\left(\frac{E_{\gamma}}{E_{b}}\right)^{-\alpha-\beta log(E_{\gamma}/E_{b})},
\label{LP}
\end{equation}
where $N_{0}$, $\alpha$ and $\beta$ are free parameters. Different models are compared using a log likelihood ratio test: the TS is twice the difference in the log likelihoods and its probability distribution can be approximated by a $\chi^2$ distribution.\\
The spectra of each source are calculated by separately running {\it gtlike} for smaller energy intervals equal in logarithmic space. In each bin the flux is estimated by fitting the PL model with the spectral slope fixed at the value obtained by the fit in the entire energy range and the bins only with $TS\geq4$ were considered.\\
Using the data accumulated for longer time than used in 3FGL can result in new \gray sources in the RoI, which are not properly accounted for in the model files. In order to probe for additional sources, a TS significance map on a square grid with $0.1^{\circ}\times0.1^{\circ}$ spacing that covers the entire RoI is created using the best-fit model. If there were excess hot-spots with  $TS > 25$ we sequentially added a new point source with a conventional spectral definition (PL) and performed binned likelihood analysis with {\it gtlike}.
\subsection{Temporal Analysis}
The \gray light curves are calculated by repeating the same analysis for shorter time periods. In the model file obtained from the whole-time analysis, the photon indexes of all background sources are fixed to the best guess values in order to reduce the uncertainties in the flux estimations but the normalization of sources within RoI are free to vary. The PL slope of each source being analyzed is first fixed and then allowed to vary in the fit. Since no variability is expected for the underlying background diffuse emission, we fixed their parameters to the average values obtained in the 7-year analysis.\\
The minimal time interval used for light curves calculation has been chosen so that the period when the source has been detected by \fermi corresponds to the large fraction of the total bins, considering the periods when TS$>4$. For some sources, pure statistics did not allow to investigate the variability in shorter than six-month bins while for bright sources the light curves with denser time sampling are considered. In order to identify whether the \gray emission is variable or not, we construct a variability index from the value of the likelihood in the null hypothesis that the source flux is constant across the considered period, and the value under the alternate hypothesis that the source is variable \citep{nolan}. If the null hypothesis is correct the variability index is distributed as $\chi^2$ with $N-1$ degrees of freedom where $N$ is the number of bins used in the light curves. The source is considered variable when the variability index is equal or greater than the threshold value of 99\% confidence level in a $\chi^2$ distribution with $N-1$ degrees of freedom (e.g., 115.88 and 16.81 for the light curves with a month and a year time-scales, respectively). This is a commonly accepted methodology used to identify variable sources in the \fermi catalogs (e.g. \citep{nolan,3fgl,acero}). In order to double check the results, possible variation is also investigated by using a simple $\chi^2$ test following \citet{var}.\\
At regular (fixed) time binning, using long bins will smooth out the fast variations during bright periods, and on the contrary, using short bins might result in many upper limits during the low-activity periods which prevents variability studies. Thus we also generate light curves with the help of the adaptive binning method \citep{lott}. Although this method has been developed for blazar studies, it can be applied to radio galaxies as well. In the adaptive binning method the bin width is adjusted by requiring a constant relative flux uncertainty, which produces longer time intervals during lower flux levels and narrower bins when the source is in a high state. This method allows to get maximum possible information about the flux variation.\\
In the adaptive binning method, the light curves are with constant uncertainties of the fluxes in each time bin. In principle, the flux can be calculated above any energy but above the optimum energy of $E_{\rm opt}$, the accumulation times necessary to fulfill the required conditions (e.g., uncertainty) are the shortest relative to other choices of energy. $E_{\rm opt}$ depends on the signal/background ratio and is calculated using the PL slope ($\Gamma$) and the flux obtained in the 0.1-300 GeV interval (see the appendix of \cite{lott}). For each source, adaptive binning light curves with 15\% and 20 \% uncertainties are generated.
\begin{table*}
\caption{Results of the Fermi-LAT observation of non-blazar AGNs from 4 August 2008 to 4 August 2015.}
\label{tab:a}
\begin{threeparttable}
\tabcolsep7pt\begin{tabular}{lcccccccc}
\hline
\hline
Name  & Class & $d_{L}$ & $\Gamma$  & $F_{\gamma}(>100\:{\rm MeV})$ & $L_{\gamma}$ & $\sigma$ & $E_{max}$ & time \\
                             &     & Mpc        &                    & $10^{-9} {\rm photon\: cm^{2}\:s^{-1}}$ & ${\rm erg\:s^{-1}}$ & & GeV & MJD \\
\hline
3C 264 & FRI & 94.6 &  $2.09\pm0.14$ & $3.48\pm1.26$  & $3.71\times10^{42}$ & 9.1 & 97.2 & 56452.9 \\ 
NGC 1275 & FRI & 76.3 &$2.10\pm0.006$ & $342.4\pm3.29$    &  $2.33\times10^{44}$ & 267.4 & 221.5 & 55402.4 \\ 
3C 120 & FRI & 145.0 &$2.73\pm0.06$ & $36.3\pm4.95$  & $3.47\times10^{43}$ & 18.3 & 4.81 & 56838.1\\ 
Cen A Core & FRI& 3.8 & $2.68\pm0.02$ & $164.9\pm2.86$ & $1.12\times10^{41}$ & 82.4 & 111.0 & 55018.5\\ 
NGC 2484 & FRI & 189.6 & $2.21\pm0.21$ & $2.49\pm1.24$    & $8.13\times10^{42}$ & 5.6 & 39.7 & 54977.4 \\ 
Cen B & FRI & 55.9 & $2.49\pm0.04$ & $65.9\pm5.09$  & $1.18\times10^{43}$ & 18.7 & 66.2 & 55044.84\\ 
M87 & FRI  & 18.4 & $2.01\pm0.04$ & $15.9\pm1.26$ & $8.00\times10^{41}$ & 32.2 & 258.8 & 55658.0 \\ 
NGC 6251 & FRI & 107.9 & $2.36\pm0.04$ & $25.3\pm1.57$ & $2.02\times10^{43}$ & 32.0 & 8.717 & 56858.0  \\ 
NGC 1218 & FRI & 125.5 & $2.04\pm0.07$ & $6.67\pm1.22$ & $1.42\times10^{43}$ & 13.5 & 14.9& 55664.4 \\ 
TXS 0331+3915 & FRI & 89.6 & $1.84\pm0.17$ & $1.49\pm0.80$ & $3.10\times10^{42}$ & 6.6 & 46.3 & 55870.7 \\ 
PKS 0625-35 & FRI & 243.7 & $1.89\pm0.04$ & $11.4\pm1.14$ & $1.48\times10^{44}$ & 30.6 & 107.9 & 57000.3\\ 
3C 111 & FRII & 215.5 & $2.76\pm0.06$ & $44.29\pm3.24$  & $9.09\times10^{43}$ & 16.8 & 6.72 & 56676.3 \\ 
3C 303& FRII  & 668.0 & $2.04\pm0.19$ & $1.44\pm0.75$   & $8.75\times10^{43}$ & 5.2 & 4.49 & 56274.9\\ 
Pictor A & FRII & 154.3 & $2.54\pm0.14$ & $12.19\pm3.36$   & $1.56\times10^{43}$ & 11.0 & 25.4 & 54948.8 \\  
3C 207 & SSRQ & 4120.2 & $2.66\pm0.09$ & $14.06\pm1.79$ & $1.15\times10^{46}$ & 11.5 & 6.0 & 54783.4\\ 
3C 221 \tnote{{\color{blue} 5}} & SSRQ &-- & $2.44\pm0.09$ & $8.64\pm1.49$  & --& 11.2 & 3.43 & 56094.1 \\ 
3C 275.1 & SSRQ & 3215.1 & $2.48\pm0.12$ & $10.03\pm2.06$ & $6.02\times10^{45}$ & 10.5 & 6.71 & 56722.4 \\ 
3C 380 & SSRQ & 4203.2 & $2.43\pm0.03$ & $37.39\pm1.88$  & $4.08\times10^{46}$ & 36.2 & 66.7 & 55463.4 \\ 
PKS 1203+04 & SSRQ & 3737.6 & $2.85\pm0.18$ & $9.89\pm2.38$ & $5.75\times10^{45}$ & 6.3 & 3.38 & 54961.0 \\ 
3C  286 & SSRQ & 5409.8 & $2.71\pm0.12$ & $11.04\pm1.85$ & $1.49\times10^{46}$ & 8.6 & 3.56 & 55038.6 \\ 
4C +39.23B & CSS &  8370.8 & $2.48\pm0.06$ & $19.11\pm1.95$  & $7.77\times10^{46}$ & 20.5 & 9.87 & 55543.0 \\ 
1H 0323+342 & NLSy1 & 273.5 & $2.86\pm0.04$ & $72.63\pm2.82$ & $2.25\times10^{44}$ & 35.9 & 3.06 & 56147.0 \\ 
SBS 0846+513 & NLSy1 & 3423.7 & $2.26\pm0.02$ & $41.78\pm1.60$ & $4.02\times10^{46}$ & 56.2 &16.5 & 56443.5 \\
PMN J0948+0022 & NLSy1 & 3426.5 & $2.64\pm0.02$ & $136.5\pm2.78$  & $7.85\times10^{46}$ & 78.3 & 4.78 & 56483.8\\ 
PKS 1502+036 & NLSy1 &  2224.4 & $2.67\pm0.04$ & $47.93\pm2.40$ & $1.13\times10^{46}$ & 31.4 & 4.62 & 56765.7\\ 
PKS 2004-447 & NLSy1 & 1204.0 & $2.63\pm0.05$ & $25.33\pm1.89$ &$1.81\times10^{45}$ & 19.1 &7.97 & 54761.7\\ 
\end{tabular}
\begin{tablenotes}
    \item[{\color{blue} 5}] No redshift found.
  \end{tablenotes}
\end{threeparttable}
\end{table*}
\begin{figure*}
  \centering
    \includegraphics[width=0.45\textwidth]{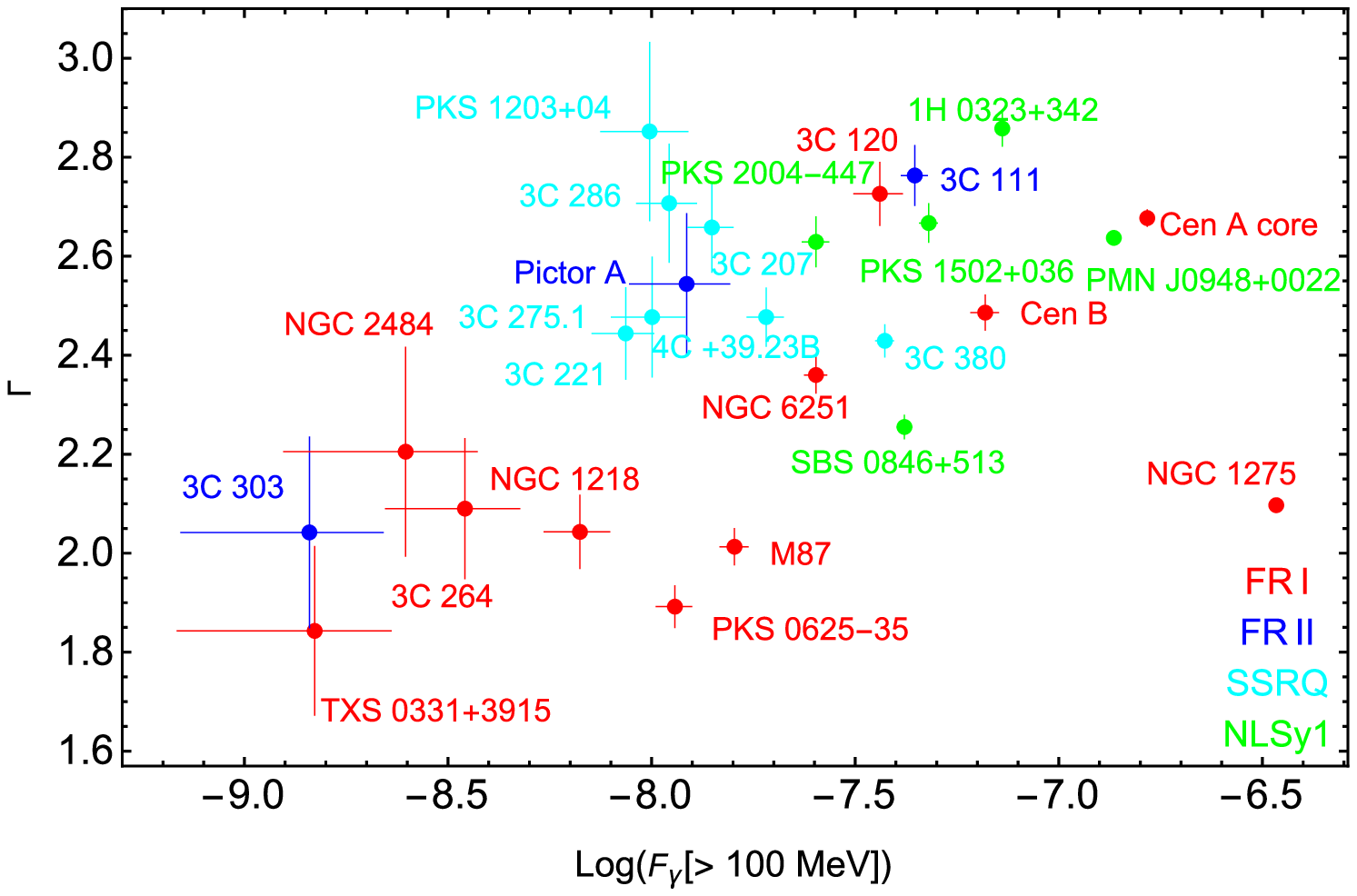}
    \includegraphics[width=0.45\textwidth]{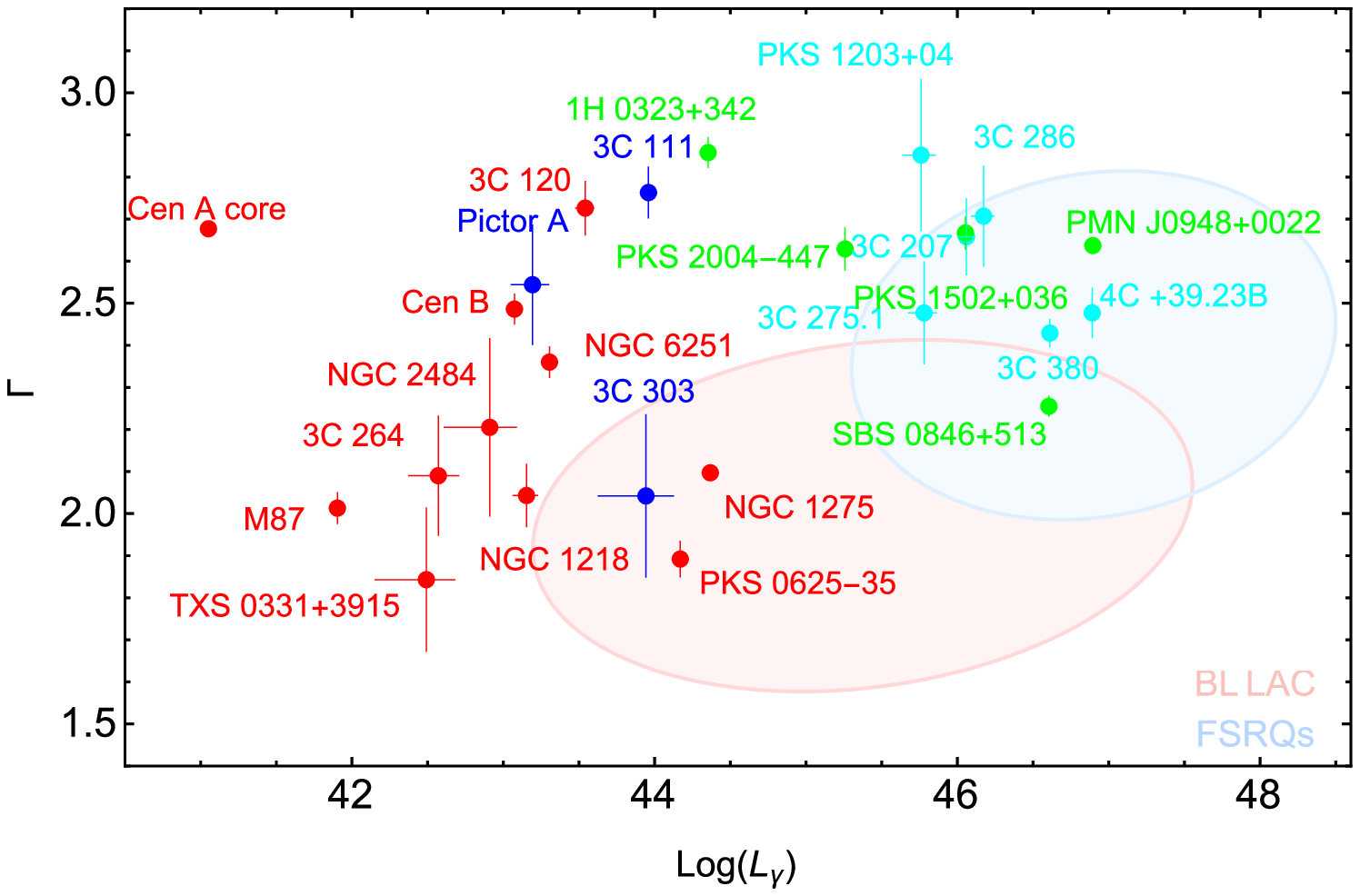}
    \caption{ (a) The flux ($E >100$ MeV) versus photon spectral index of FRIs, FRIIs, SSRQs and NLSy1s. (b) Photon spectral index vs. \gray luminosity of the considered sources as compared with similar properties of blazars (light blue ellipse for FSRQs and light red for BL Lacs). In both plots FRIs, FRIIs, SSRQs and NLSY1s are shown with red, blue, cyan and green colors, respectively.}%
    \label{all}
\end{figure*}
\section{Results}\label{sec2.2}
Of the 27 non-blazar AGNs studied here, 26 are detected with TS$>$ 25. The only exception is TXS 0348+013 which appeared with TS$\simeq$14. The results from the current data analysis are summarized in Table \ref{tab:a}. For each source, the class, distance, \gray photon index ($\Gamma$), flux ($F_{\gamma}$) and luminosity ($L_{\gamma}$), the detection significance ($\sigma=\sqrt{TS}$) and the energy of the highest-energy events with their detection times are presented. Our results have been verified by {\it i)} carrying out analyses by two different ways: first by a standard procedure as suggested by the \fermi collaboration \footnote{https://fermi.gsfc.nasa.gov/ssc/data/analysis/scitools/} and then by using the {\it Fermipy} python package \footnote{http://fermipy.readthedocs.io/en/latest/}, and {\it ii)} by comparing the results with the previously reported ones (from 3FGL).\\
The results presented here are consistent with those reported in \citet{3fgl} but with reduced uncertainties and a higher detection significance. The majority of the sources have been detected with $\geq10\sigma$ which allowed to perform spectral studies above 1 GeV as well. NGC 1275 was detected with the highest detection significance of $267.4\sigma$ while 3C 303 and NGC 2484 are still detected at the threshold limit with $5.2\sigma$ and $5.6\sigma$, respectively, because these sources  were mostly undetected by \fermi during the last years of the considered period.\\
From Table \ref{tab:a}, the \gray photon index of non-blazar AGNs varies from $\Gamma=1.84$ to $2.86$. The hardest photon indexes of $\Gamma\simeq1.84\pm0.17$ and $\Gamma\simeq1.89\pm0.04$ have been found for TXS 0331+3915 and PKS 0625-35, respectively, while the softest spectrum is observed for 1H 0323+342, $\Gamma\simeq2.86\pm0.04$. Interestingly, 3C 264, NGC 1275, M87, NGC 1218 and 3C 303 have been detected with $\Gamma\simeq2.0-2.1$ which corresponds to a nearly flat spectrum in the $\nu F_{\nu}({\rm erg\:s^{-1}})$ representation. The brightest sources, Cen A and NGC 1275, have a flux of $\geq10^{-7}\:{\rm photon\:cm^{-2}\:s^{-1}}$ and the faintest source 3C 303- $(1.44\pm0.75)\times\:10^{-9}\:{\rm photon\:cm^{-2}\:s^{-1}}$.\\
Among FRIs, PKS 0625-35 ($d_{L}\approx243.7$ Mpc) is the most distant source detected so far while the other FRIs are mostly at $d_{L}\leq200$ Mpc. Except for Cen A and NGC 1275, which are bright due to their proximity, as the flux of FRIs is relatively weak (a few times exceeding the \fermi sensitivity of $\approx1.9\times10^{-10}\;{\rm photon\:cm^{-2}\:s^{-1}}$ at 10 GeV), it is not expected to detect FRIs that are much farther than those currently being observed. On the contrary, the luminous FRIIs, SSRQs (CSS) and NLSy1s are detected at much larger distances - $z>0.1$. For instance, the most distant non-blazar AGNs detected in the \gray band are 3C 286 and 4C +39.23B at $z=0.85$ and $z=1.21$, respectively. If the observed luminosities are typical for FRIIs, SSRQs (CSS) and NLSy1s, in principle more distant sources can be still detected (e.g., the \gray emission from 4C +39.23B still would be detected by \fermi even if the source is at $z\simeq1.84$). NLSY1s detected in the \gray band are at $z\simeq(0.2-0.6)$ (except the nearby 1H 0323+342 at $z=0.061$), which is not surprising, considering their powerful jets.\\
The results presented in Table \ref{tab:a} are shown in Fig. \ref{all}, the left panel of which shows the \gray photon index ($\Gamma$) versus flux ($F_{\gamma}[>\:100\:{\rm MeV}]$) estimated from the PL fit. It is evident that the considered sources have different properties as they are well separated in the $\Gamma-F_\gamma$ plane. For example, FRIs have a harder \gray photon index, the mean being $\Gamma_{\rm m}\simeq2.22$, and a \gray flux ranging from $\simeq1.5\times10^{-9}\times{\rm photon\:cm^{-2}s^{-1}}$ to $\simeq3.6\times10^{-8}\times{\rm photon\:cm^{-2}s^{-1}}$. Among FRIs, only Cen A and NGC 1275 are showing systematically different properties. In case of Cen A, the photon index and the flux are mostly defined by the low energy component (see the next Section), but when using those of the second component ($\Gamma=2.31\pm0.1$ and $(9.16\pm0.68)\times10^{-10}\;{\rm photon\:cm^{-2}\:s^{-1}}$), it fits well in the common properties of FRIs. NGC 1275 is a unique and particularly bright radio galaxy, so it is not a surprise that it demonstrates quite different properties. 
The FRIIs/SSRQs have rather similar properties and occupy a narrow area on the $\Gamma-F_{\gamma}$ diagram: the \gray photon index is $(2.3-2.7)$ with the mean being $\Gamma_{\rm m}\simeq2.58$ and a flux of $(0.9-4)\times10^{-8}\:{\rm photon\:cm^{-2}s^{-1}}$. Only 3C 303 with $\Gamma=2.04\pm0.19$ is out of the mentioned area, showing properties more characteristic of FRIs. Despite using the data having been accumulated for seven years, this source is detected only with $5.2\sigma$, causing large uncertainties in the parameter estimation. For NLSy1s $\Gamma>2.3$ which is more similar to FRIIs/SSRQs albeit their \gray flux is shifted to $\geq10^{-7} \times {\rm photon\:cm^{-2}s^{-1}}$. Worth of mentioning is the bright source PMN J0948+0022 which, being located at a distance of 3.4 Gpc, appears with the same flux as does the nearby Cen A (3.8 Mpc) but its luminosity is $\sim7\times10^5$ times higher.\\
Even if the non-blazar AGNs considered here may have different jet properties (e.g., jet viewing angle, superluminal apparent motion etc.), it is interesting to compare their \gray photon index and luminosity with those of blazars. The right panel of Fig. \ref{all} shows the \gray photon index versus luminosity of  non-blazar AGNs and of those BL Lacs and FSRQs  which are included in the third catalog of AGNs detected by \fermi \citep{3fgl}. The \gray luminosity is estimated from the measured flux by
\begin{equation}
L_{\gamma}=4\:\pi d_{L}^2\:E_{\rm max}\left(\frac{(E_{\rm min}/E_{\rm max})^{2-\Gamma}-1)}{(E_{\rm min}/E_{\rm max})^{1-\Gamma}-1}\right)\:F_{\gamma}(>\:{\rm 100\: MeV})
\label{Lum}
\end{equation}
where $E_{\rm min}=100\:{\rm MeV}$ and $E_{\rm max}=300\:{\rm GeV}$. The luminosity distance ($d_{L}$) is calculated by adopting the standard cosmological parameters of $H_{0}=70\:{\rm km s^{-1}Mpc^{-1}}$, $\Omega_{m}=0.27$ and $\Omega_{\Lambda}=0.73$.\\
\begin{figure*}
\includegraphics[width=0.45\textwidth]{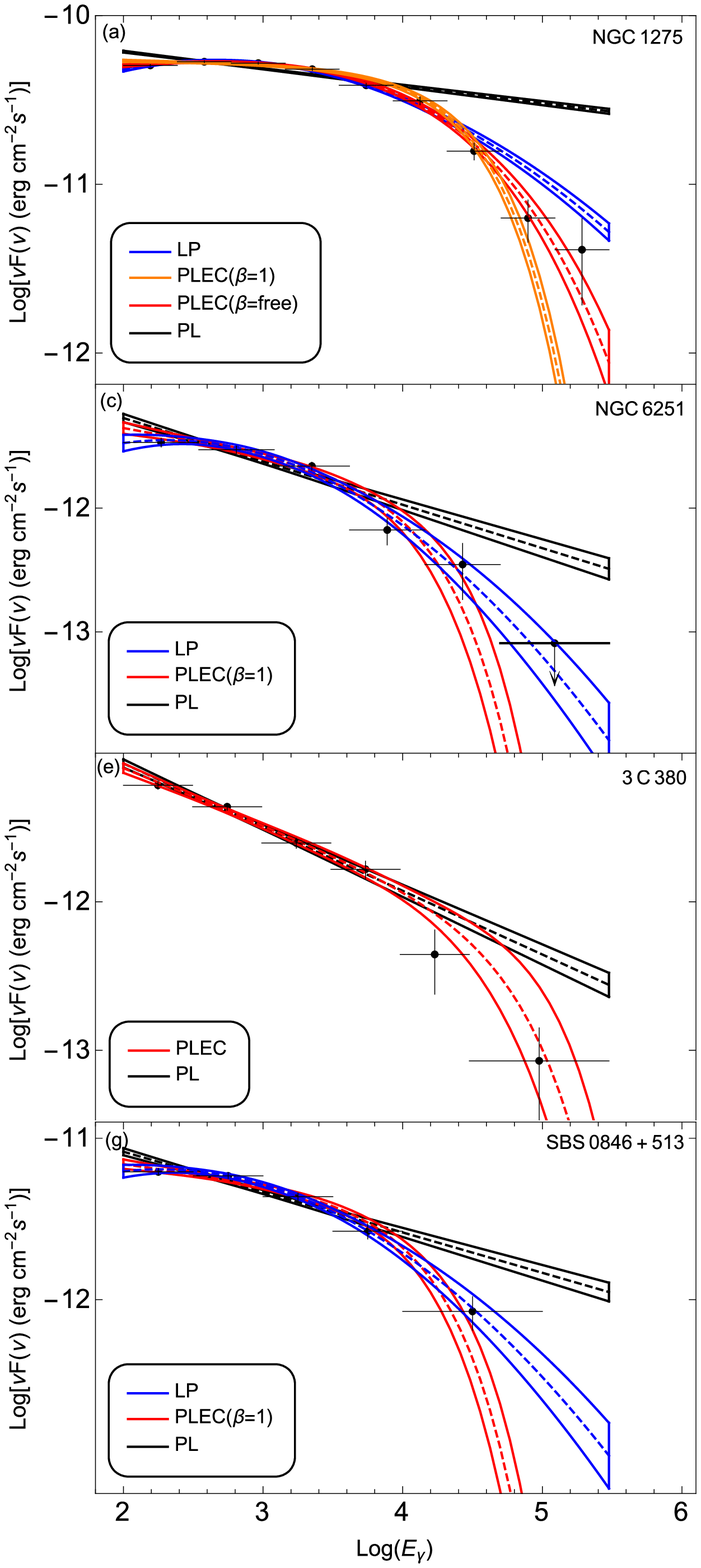}
\includegraphics[width=0.45\textwidth]{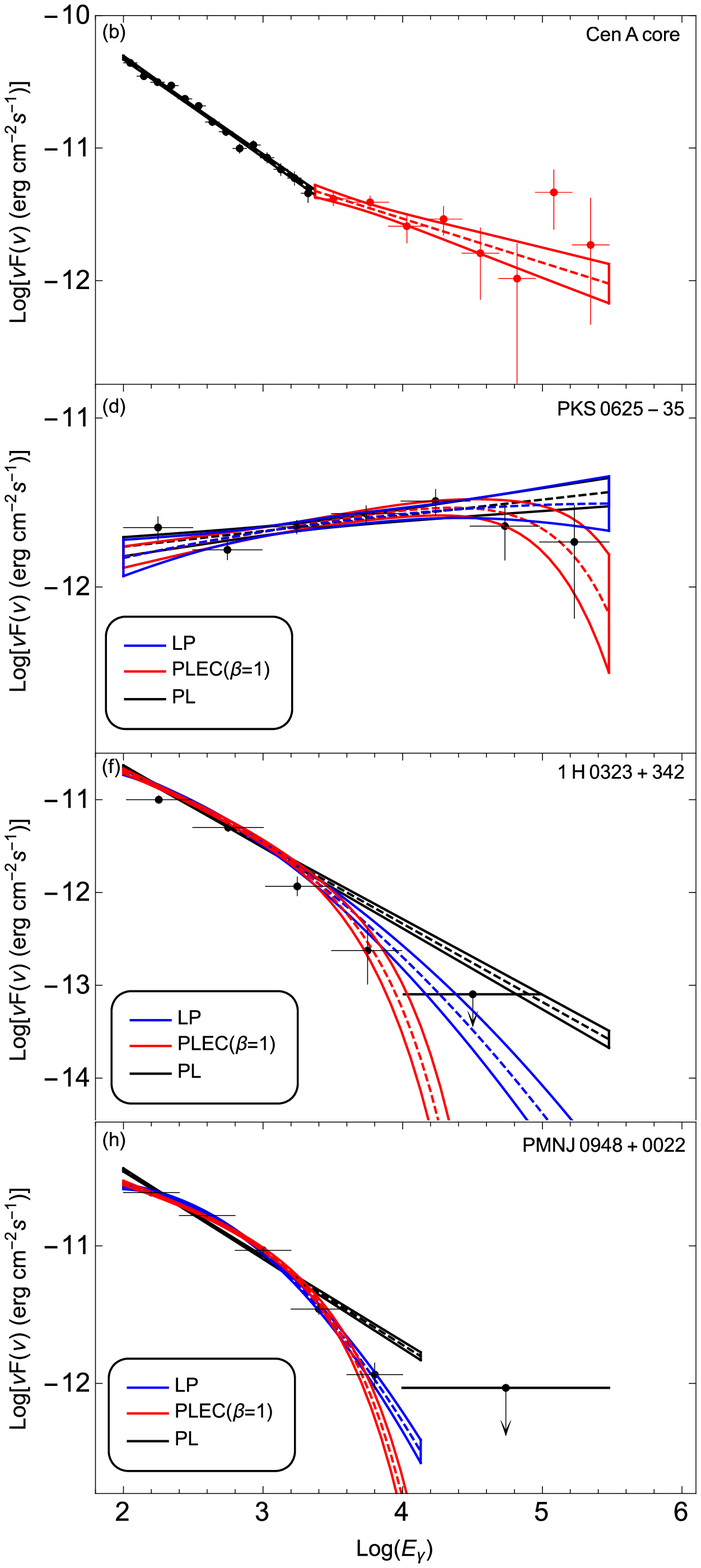}\\
   \caption{Spectra of NGC 1275, Cen A core, NGC 6251, PKS 0625-35, 3C 380, SBS 0846+513, 1H 0323+342 and PMN J0948+0022 measured in the energy range from 100 MeV to 300 GeV, which showed deviation from the simple PL model above several GeV (see the text). The best fit models obtained with {\it gtlike} are depicted together with spectral points which have been obtained by separately running {\it gtlike} for smaller energy intervals. The corresponding parameters are given in Table \ref{tab_SD}.}
   \label{sed_all}
\end{figure*}
The previous studies of non-blazar AGNs reveal that in the $\Gamma-L_{\gamma}$ plane FRIs and FRIIs occupy different regions as do BL Lacs and FSRQs \citep{abdomisaligned}.
 In case of blazars, it has been already shown that there is a well-defined boundary between BL Lacs and FSRQs. BL Lacs appear with $\Gamma<2.2$ and $L_{\gamma}<\:10^{47}\;{\rm erg\:s^{-1}}$, whereas FSRQs with $\Gamma>2.2$ and $L_{\gamma}>\:10^{47}\;{\rm erg\:s^{-1}}$; the two blazar subclasses overlap at intermediate values of both spectral index and luminosity \citep{ghisellini09}. It has been suggested that this division is of a pure physical origin, i.e. in BL Lacs the jet propagates in a medium where the external radiation field is absent or very weak, so that the electrons are cooling less but can be accelerated to VHEs, so that low luminosity and a harder electron/photon index are expected. In FSRQs, the emitting electrons will efficiently cool down and reach only moderate energies, therefore they will appear as luminous sources, but with a soft \gray photon index. Interestingly, similar division is noticeable also for FRIs and FRIIs, albeit shifted to lower luminosities (Fig. \ref{all}, right panel; the blazars have higher luminosity due to Doppler boosting). The \gray luminosity of most of FRIs is $L_{\gamma}\leq\:3\times10^{44}\;{\rm erg\:s^{-1}}$ and the photon index  $\Gamma\leq2.3$. Likewise, for FRIIs $L_{\gamma}\geq 3\times\:10^{44}\;{\rm erg\:s^{-1}}$ and $\Gamma>2.3$. SSRQs and NLSy1s occupy the area beyond $\Gamma>2.3$ and $L_{\gamma}>\:10^{45}\;{\rm erg\:s^{-1}}$ and are clearly separated from FRIs and FRIIs, except for 1H 0323+342, the nearest NLSy1 having been detected so far in the \gray band, which has a lower luminosity of $L_{\gamma}<\:10^{45}\;{\rm erg\:s^{-1}}$. The difference of \gray luminosity in different source classes is in a good agreement with the unification theories: less luminous FRIs are followed by FRIIs and then by SSRQs - most aligned version of FRIIs.\\
For the sources studied here it is interesting to compare the energy released in the \gray band with the capability of the central source (Eddington accretion power). More reasonable would be to compare the \gray luminosity in the proper frame of the jet ($L_{em,\gamma}\simeq L_{\gamma}/2\Gamma_{\rm bulk}^2$) but it requires to make different assumptions to estimate unknown $\Gamma_{\rm bulk}$. Considering that $\Gamma_{\rm bulk}$ is usually not very high for non-blazar AGNs, it is reasonable to use $L_{\gamma}$. We did not find black hole masses for Cen B, 3C 221, PKS 1203+04, TXS 0331+3915 and 4C +39.23B while for other sources the values available in the literature have been used. A common feature for all FRIs is that the intrinsic \gray luminosity corresponds to only a small portion of the Eddington accretion power - $L_{\gamma}/L_{\rm Edd}\leq10^{-4}$. For example, M87 hosts a super massive black hole with an enormous mass of $(6.6\pm0.4)\times10^{9}\:M_{\odot}$ \citep{gebhard11} but $L_{\gamma}/L_{\rm Edd}\simeq9.6\times10^{-7}$ implying that in the \gray band only a very small fraction of the source capability is realized. Similarly, $L_{\gamma}/L_{\rm Edd}\simeq3.0\times10^{-5}$ for the nearby Cen A and $L_{\gamma}/L_{\rm Edd}\simeq9.1\times10^{-4}$ for the distant PKS 0625-35. The only exceptions are 3C 120 and NGC 1275, where $L_{\gamma}/L_{\rm Edd}\simeq5\times10^{-3}$ which is not surprising considering these are the unique FRIs, where the \gray emission is characterized by rapid changes. For FRIIs and SSRQs $L_{\gamma}/L_{\rm Edd}=10^{-3}-10^{-4}$ and $0.2-0.4$, thus the total available power is very efficiently converted to the emission in the \gray band. Interestingly, due to the small black hole masses in NLSY1s, $L_{\gamma}/L_{\rm Edd}=(2-4)$ which can be even higher during the flaring periods, similar to blazars,  where also such "super-Eddington" regimes (but with larger $L_{\gamma}/L_{\rm Edd}$ ratio) are observed [e.g., \citep{bonnoli}].\\
The highest energy of the events with higher probability of being associated with the sources and having less angular separation and their detection times, computed using the {\it gtsrcprob} tool, are presented in Table \ref{tab:a}. The highest-energy photons of $258.8$, $221.5$, $111.0$ and $107.9$ GeV were detected from M87, NGC 1275, Cen A and PKS 0625-35, respectively. Except Cen A, all the other sources have comparably hard photon indexes, so their spectra extend to HEs, allowing to detect photons above $100$ GeV. Moreover, when the highest energy photons from M87 and PKS 0625-35 were detected their \gray spectra were unusually hard, $1.50\pm0.20$ and $1.74\pm0.18$, respectively. For the other sources, due to the soft PL indexes the photon energies are limited to $<50$ GeV; the only exception is 3C 380 from which photons with $66.7$ GeV have been detected. Even though NLSY1s are powerful \gray sources, due to the break in their spectra around several GeV, the detected photon energies are mostly below $10$ GeV, except for 1H 0323+342 with 16.5 GeV.
\section{Discussion}\label{sec3}
\begin{table*}
\scriptsize
\caption{The parameters of fitting with LP or PLEC models.}
\centering
\begin{tabular}{l c c c c c c c } \hline
Name & Model  & $\alpha$ & $\beta$ & ${\rm E_{\rm cut}}$&$F_{100}$ & $2(\Delta\mathcal{L})$\\
   & & & & (GeV)& (10$^{-8}$ ph cm$^{-2}$ s$^{-1}$)& &  \\\hline
NGC 1275 & PLEC & $1.86\pm0.03$ & $0.43\pm0.04$ & $6.24\pm2.30$ & $31.84\pm0.38$ &34.1\\
NGC 6251 & LP & $2.29\pm0.05$ & $0.1\pm0.03$ &-- & $2.10\pm0.19$ & 16.2\\
PKS 0625-354 & PLEC & $1.83\pm0.06$ & 1 & $131.2\pm88.04$ & $1.04\pm0.13$ & 4.0  \\
3C 380 & PLEC & $2.36\pm0.05$ & 1 & $55.57\pm50.74$ & $3.21\pm0.19$ & 2.0  \\
1H 0323+342 & LP & $2.74\pm0.06$ & $0.09\pm0.04$ & -- & $7.03\pm0.30$ & 9.6  \\
SBS 0846+513 & LP & $2.15\pm0.03$ &$0.09\pm0.02$ & -- & $3.21\pm0.18$& 25.8  \\
PMN J0948+0022 & LP & $2.41\pm0.02$ & $0.16\pm0.01$ & -- & $12.75\pm0.02$ & 95.2  \\\hline
\end{tabular}
\label{tab_SD}
\end{table*}
\begin{figure*}
\hspace*{-1.5cm}
  \centering
    \includegraphics[width=0.96\textwidth]{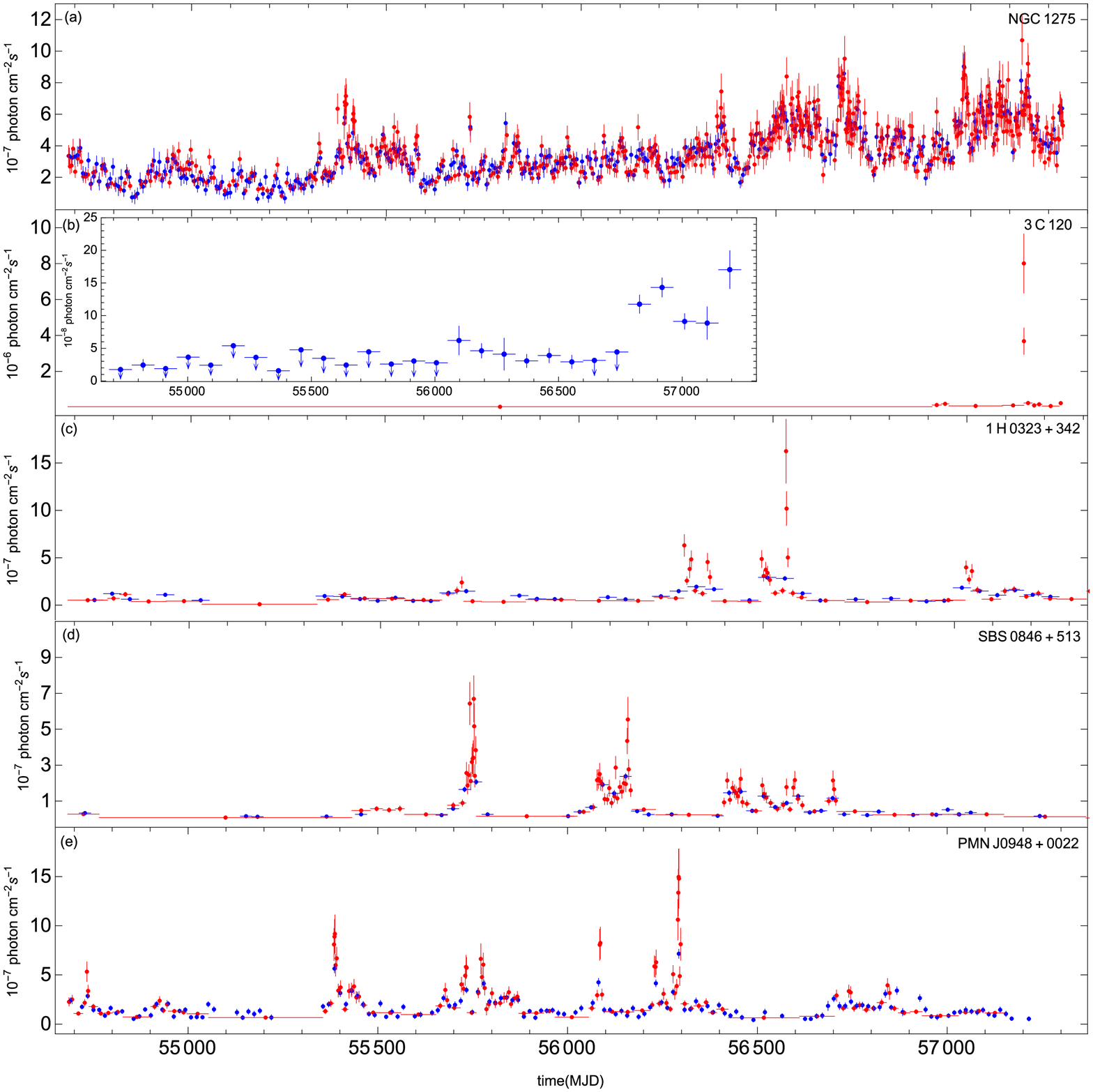}
  \caption{ Light curves of NGC 1275, 3C 120, 1H 0323+342, SBS 0846+513 and PMN J0948+0022. The light curves are calculated with standard {\it gtlike} tool for normal (blue)  and adaptively binned (red) time scales.}
\label{var_all}
\end{figure*}
The \gray spectra of most of the sources studied here are described significantly better by a PL. However, the \gray spectra of some bright sources having been detected with a higher significance show a tendency of a cutoff/break above 1 GeV. In addition, non-blazar AGNs variability studies can shed light on the physical processes in action, such as particle acceleration and emission mechanisms, origin of brightening(s) and the structure and location of the emission region(s). Below, the non-blazar AGNs detected by \fermi are discussed individually or by each set of objects that behave in a similar fashion. The spectra and light curves are presented in Figs. \ref{sed_all} and \ref{var_all} and the main parameters are in Tables \ref{all} and \ref{tab_SD}.\\
\textit{NGC 1275:} The \gray spectrum is shown in Fig. \ref{sed_all} (a). The spectrum starts to steepen around $\simeq(1-10)\;{\rm GeV}$ and more complex LP and PLEC models are used to fit the data. The log likelihood ratio test shows that the PLEC modeling when $\beta=0.43\pm0.04$ is preferred over the simple PL with a significance of $5.8\sigma$. The PLEC model is shown with red color in Fig. \ref{sed_all} (a) together with LP (blue) and PLEC ($\beta=1$, orange) models.\\
Among the considered sources, NGC 1275 is unique, where the flux variation in time scales as short as a day can be investigated. Recently, we showed that during the major \gray flares of NGC 1275 in December 2016/January 2017, the source reached a daily peak flux of $(2.21\pm0.26)\times10^{-6}\;{\rm photon\:cm^{-2}\:s^{-1}}$ and had an e-folding time as short as $1.21\pm0.22$ hours. Even if it was known that the \gray emission from this source was varying in month \citep{abdongc1275,kataokangc1275} and sub-week time scales \citep{brown}, such rapid \gray flux variations had never been previously observed from any radio galaxy in the \gray band. The light curve with normal (7-day) and adaptive time binning (assuming 20 \% uncertainty) is shown in Fig. \ref{var_all} (a) with blue and red colors, respectively. The \gray light curve appears to be quite a complex one, with many peaks and flaring periods. The adaptively binned light curve shows that in the active states the required uncertainty is reached on sub-day scales, which once again confirms the short-scale variability of this source. Also, after MJD 56000  (2012) the source is mostly in a bright state which resulted in the increase of the average flux level (Fig. \ref{var_all} [a]).\\
\textit{3C 120:} The light curve with 90-day binning (see the insert in Fig. \ref{var_all} [b]) shows that up to MJD 56900 the source flux is below the \fermi sensitivity threshold and is mostly undetectable (only upper limits).  Afterwards, the \gray flux increased being detected by \fermi almost all the time.  More dramatic changes in shorter time scales can be seen on the adaptively binned light curve shown in Fig. \ref{var_all} (b) (red data). The data collected from MJD 54682 (4 August 2008) up to MJD 56919.31(19 September 2014) are necessary to reach the 20\% uncertainty in the first bin, while it took shorter time afterwards. The most dramatic increase in the \gray flux was observed on April 24, 2015 when within 19.0 min the flux reached $(7.46\pm1.56)\times10^{-6}\;{\rm photon\:cm^{-2}\:s^{-1}}$ with $\Gamma=2.29\pm0.21$ and $11.2\sigma$ detection significance. Then the flux slowly decreased and the data collected within 10 to 35 days were enough to reach 20 \% uncertainty. A more detailed analysis of the origin of emission using multiwavelength data is presented in \citet{zargaryan17}.\\
\textit{Cen A core:} The changes observed in the \gray spectrum are shown in Fig. \ref{sed_all} [b]. The presence of the unusual break \citep{sahakyancen,browncena} is confirmed using the \fermi data accumulated for a longer period. In the energy range of 0.1-300 GeV, the broken PL with the break energy $2.35\pm0.08$ GeV is preferred over the PL at $\simeq5.0 \sigma$ level. The energy range of 0.1-300 GeV was divided into (0.10-2.35) GeV and (2.35-300) GeV subintervals and the likelihood analyses were re-run. The first component with $\Gamma=2.75\pm0.02$ and a photon flux of $(1.62\pm0.04)\times10^{-7}\;{\rm photon\:cm^{-2}\:s^{-1}}$ is detected with $71.5\sigma$ and shown with black color in Fig. \ref{sed_all} (b) while the red line shows the second component with $\Gamma=2.31\pm0.1$ and $(9.16\pm0.68)\times10^{-10}\;{\rm photon\:cm^{-2}\:s^{-1}}$ detected with $25.7\sigma$. This break is unusual in that the spectrum gets harder instead of getting softer, while typically the opposite occurs. This is a unique case and has never been yet observed in the \gray spectra of other AGNs. This hardening could be caused by the contribution of an additional HE component beyond the common SSC jet emission.\\
\textit{NGC 6251:} Detected with $32.0\:\sigma$, its spectrum extends up to $\simeq(60-70)$ GeV, but the PL model fails to explain the \gray data above 1 GeV. The results of fitting with alternative models, LP (blue) and PLEC ($\beta=1$, red), are shown in Fig. \ref{sed_all} (c). The data are better explained by LP with $\alpha=2.29\pm0.05$ and $\beta=0.1\pm0.03$. Among the considered radio galaxies, only the spectrum of NGC 6251 is described by LP, indicating that perhaps a different process is responsible for the particle acceleration. As shown by \citet{massaro}, the \gray spectra well described by the LP are formed when the jet electrons are accelerated stochastically.  Instead, the PLEC is formed within diffuse shock acceleration theories and can be interpreted as the emission from accelerated particles, where the cutoff is due to {\it i)} escape of the highest energy particles, {\it ii)} limitation of the acceleration processes or {\it iii)} radiative losses.\\
\textit{PKS 0625-25:} This is one of the radio galaxies detected in the VHE \gray band \citep{dyrda}. Its \gray photon index, $\Gamma=1.89\pm0.04$, is unusually harder as compared with those of other radio galaxies. Around $100$ GeV, there is a hint of substantial deviation from PL (black curve in Fig. \ref{sed_all} [d]), but the PLEC model with $\Gamma=1.83\pm0.06$ and ${\rm E}_{\rm cut}=131.2\pm88.04$ GeV (Fig. \ref{sed_all} [d], red curve) is preferred over the simple PL only at the $\sim2 \sigma$ level, so that the PL can not be rejected. However, the spectrum of PKS 0625-35 in the TeV band is softer, $\Gamma_{\rm TeV}=2.8\pm0.5$ \citep{dyrda}, which strengthens the assumptions of the cutoff in the GeV band.\\
\textit{3C 380:} Along with 3C 207, it is the first SSRQ detected by \fermi in the \gray band \citep{abdomisaligned}. Its \gray spectra can be described by PL with $\Gamma=2.43\pm0.03$ detected with $\approx36.2\sigma$ and its emission extends up to 300 GeV (Fig. \ref{sed_all} [e]). We note an indication of deviation of the PL with respect to the data above 10s of GeV but the PLEC modeling with $\Gamma=2.36\pm0.05$ and $E_{\rm c}=(55.57\pm50.74)$ GeV (Fig. \ref{sed_all} (e) red color) is preferred over PL only at the level of $\sim1.4\sigma$. Its \gray emission is variable in 30-day scales but the adaptively binned light curve shows that sometimes the flux increases also in short periods: for example on MJD 55184.71 within 21.3 days the flux increased up to $(8.45\pm1.37)\times10^{-8}\;{\rm photon\:cm^{-2}\:s^{-1}}$.\\
Noticeably different spectral and temporal properties are observed for NLSY1s analyzed here. The spectra of PKS 1502+036 and PKS 2004-447 were better described by PL with steep, $\Gamma=2.67\pm0.04$ and $\Gamma=2.63\pm0.05$ photon indexes, respectively. Their emission extends up to $10-30$ GeV and is variable in long (e.g., 45- and 60-day) time scales. Instead, the \gray emission of 1H 0323+342, SBS 0846+513 and PMN J0948+0022 significantly increases in day- or a few-days scales \citep{abdoNLSY,Paliya,dammando2012, dammando2013,abdopmnj, foschini}. With the help of the light curves generated by the adaptive binning method, we report on flux variations and spectral changes in shorter time scales.\\
\textit{1H 0323+342:} Its spectrum is steep, $\Gamma=2.86\pm0.04$, and the \grays are detected up to 10 GeV. There is a marginal evidence of a spectral curvature around 1 GeV (Fig. \ref{sed_all} (f) black curve) and both LP or PLEC ($\beta=1$) models can better explain the observed data (blue and red curves in Fig. \ref{sed_all} (f), respectively). However, because of low statistics ($\sim3.1 \sigma$) PL modeling cannot be excluded. For most of the time, the \gray flux of 1H 0323+342 did not much exceed the value averaged over the whole observational time (blue data in Fig. \ref{var_all} [c]). The adaptively binned light curve calculated for 15\% uncertainty shows several periods when the flux significantly increased (Fig. \ref{var_all} [c] red data). The most significant increase was observed on MJD 56534.3 when within 16.2 hrs the flux was as high as $(1.98\pm0.31)\times10^{-6}\;{\rm photon\:cm^{-2}\:s^{-1}}$ and the photon index hardened to $\Gamma=2.22\pm0.15$ .  During this flare, the flux-doubling time was as short as 0.49 days, implying the emission is produced in a very compact region, $\simeq1.2\times10^{15}\times\delta$ cm \citep{bagh}. After this rapid flare, the source again entered its quiescent state.\\
\textit{SBS 0846+513:} Below $\simeq10$ GeV its spectrum can be well described by a PL (black curve in Fig. \ref{sed_all}  [g]) with $\Gamma=2.26\pm0.02$ which is the hardest photon index obtained for the selected NLSY1s. At higher energies, the PL model overpredicts the observed data (blue line in Fig. \ref{sed_all}  [g]) suggesting a turn-over in the spectrum. Indeed, at $5.1\sigma$ the LP with $\alpha=2.15\pm0.03$ and $\beta=0.09\pm0.02$ explains the data statistically better. A sequence of flares can be identified on the \gray light curve of SBS 0846+51 with 30-day (blue) and adaptive bins (red, 20\%) shown in Fig. \ref{var_all} [d], though with a lower amplitude as compared with the flares of 1H 0323+342. The highest flux of $(6.35\pm1.12)\times10^{-7}\;{\rm photon\:cm^{-2}\:s^{-1}}$ was observed on MJD 56145.6 and the active state lasted till MJD 56146.8, when within 19.6 hours the flux was $(4.15\pm1.04)\times10^{-7}\;{\rm photon\:cm^{-2}\:s^{-1}}$, but the photon index was significantly hard $1.73\pm0.14$. This is the hardest \gray photon index detected during the considered seven years and, interestingly, the highest-energy 16.5 GeV photon had been also observed during this period. This hardening is most likely associated with the emission from the re-accelerated or fresh electrons which produced also the observed highest-energy photon.\\
\textit{PMN J0948+0022:} Its spectrum measured in the narrow 0.1-10 GeV range appeared with a significant curvature Fig. \ref{sed_all} (h). The comparison of LP and PL models yields a log likelihood ratio of $95.2$, so the PL modeling can be rejected with a significance of $9.76\sigma$. The LP model compared with the observed spectrum is shown in Fig. \ref{sed_all} (h) with blue color (for parameters see Table \ref{tab_SD}). The \gray light curve with 15-day (blue) and adaptive bins (red, 20\%) shows that the source is alternatingly in its flaring or quiescent states (Fig. \ref{var_all} [e]). Several times, the source flux was above $10^{-6}\;{\rm photon\:cm^{-2}\:s^{-1}}$ and during an extremely active state from MJD 56290.96 to MJD 56294.60, it reached a peak flux of $(1.51\pm0.31)\times10^{-6}\;{\rm photon\:cm^{-2}\:s^{-1}}$ within 18.9 hrs, corresponding to an apparent isotropic \gray luminosity of $L_{\gamma}\approx1.09\times10^{48}\;{\rm erg\;s^{-1}}$, which is typical for the brightest FSRQs. For most of the time the photon index is $>2.5$, the hardest one being $\Gamma=2.13\pm0.18$. The observed rapid flux changes in intra-day scales and the amplitude of the increase once more show that in the \gray band NLSY1s behave similar to blazars.\\
\textit{Sources with long time scale variability:} The \gray emission from some sources varies within longer time periods as compared with NGC 1275 and 3C 120. For example, the low energy component of Cen A core emission is variable within 45 days which confirms the marginal evidence reported in \citep{sahakyancen}. In the second emission component, there is an indication (at the limit) of a year and half-year scale variability, but because of limited statistics, however, no definite conclusion can be drawn. There is a hint of variability in the \gray light curves of M87 and PKS 0625-35 with 60-day binning but the statistics does not allow neither to confirm nor explore it in shorter time scales. The previous observations of M87 in the HE and VHE \gray bands have already suggested that it is in its quiescent state \citep{abdom87,abramowski} while being active and repeatedly in the flaring state in 2005 \citep{aharm87,albert,acciari,levinson}. Interestingly, these sources are also the ones which have been detected in the VHE \gray band so far \citep{rieger17}.\\ 
Among FRII/SSRQs/CSS, variable are 3C 111, 3C 207, 3C 275.1, 3C 380 and 4C+39.23B. The variability of 3C 111 is highly significant in the light curve with 60-day binning. For 3C 207 and 3C 275.1, the significance is still not high enough to confirm the variability in short time scales but the \gray emission is variable within a year. For 4C +39.23B, in the light curves with 90-day and 60-day binning, the periods when the source was and was not detected by \fermi are comparable, therefore no definite conclusion can be made but the \gray emission is variable within a half-year period. Nevertheless,  a short time scale activity around 21 April 2011 was observed, when within 29.4 days the flux was $(4.39\pm0.85)\times10^{-8}\;{\rm photon\:cm^{-2}\:s^{-1}}$ (as inferred from the adaptively binned light curve).\\
The significance is still not high enough to search for \gray variability in the other sources studied here. It is interesting to highlight Cen B and NGC 6251 where the \gray emission is detected with high $18.5\sigma$ and $31.6\sigma$ significances, respectively. The constant flux assumption can not be rejected when using light curves with short (few months) or long times scales (half-year/year). Although this does not allow to rule out possible variability completely, this is interesting, since both have large scale structures, $\sim 380$ kpc for Cen B \citep{jones} and $\sim1.8$ Mpc for NGC 6251 \citep{waggett}, and in principle, the \grays can be produced in these regions.
\section{Conclussions}\label{sec4}
The \gray emission from non-blazar AGNs included in the third catalog of AGNs detected by \fermi is investigated using the \fermi data accumulated in 2008-2015.
The sample consists of eleven FRI, three FRII, six SSRQ, one CSS and five NLSy1 sources. We obtained the following results:
\begin{itemize}
     \item[{\it i)}] The \gray photon index of non-blazar AGNs changes within $1.84-2.86$. The hardest photon indexes of $\Gamma=1.84\pm0.17$ and $\Gamma=1.89\pm0.04$ were obtained for TXS 0331+3915 and PKS 0625-35, respectively, and the softest for 1H 0323+342, $\Gamma=2.86\pm0.04$. The measured faintest flux is of the order of a few times $10^{-9}\:{\rm photon\:cm^{-2}s^{-1}}$, while the flux from bright sources exceeds $10^{-7}\:{\rm photon\:cm^{-2}s^{-1}}$. The farthest non-blazar AGN detected by \fermi is 4C +39.23B at $z\simeq1.21$ with a \gray flux of $(1.91\pm0.19)\times 10^{-8}\:{\rm photon\:cm^{-2}s^{-1}}$.
     
     \item[{\it ii)}] In the $\Gamma-F_{\gamma}$ plane, the FRIs have a harder \gray photon index with $\Gamma_{\rm m}\simeq2.22$ and a flux ranging from $\simeq1.5\times10^{-9}\times{\rm photon\:cm^{-2}s^{-1}}$ to $3.6\times10^{-8}\times{\rm photon\:cm^{-2}s^{-1}}$. The mean photon index of FRIIs/SSRQs shifts to a higher value of $\Gamma_{\rm m}\simeq2.58$ (i.e., softer spectra) and the flux to the range of $(0.9-4)\times10^{-8}\:{\rm photon\:cm^{-2}s^{-1}}$. With a \gray flux of $\geq10^{-7} \times {\rm photon\:cm^{-2}s^{-1}}$, the average photon index of NLSYs  is similar to that of FRIIs/SSRQs with $\Gamma>2.3$.
     
     \item[{\it iii)}] In the $\Gamma-L_{\gamma}$ plane we confirm the previous results from the analysis of a smaller sample of non-blazar AGNs. The \gray luminosity of most of FRIs is $L_{\gamma}<\:3\times10^{44}\;{\: \rm erg\:s^{-1}}$ with $\Gamma<2.3$ while for FRII $L_{\gamma}>3\times10^{44}\:{\rm erg\:s^{-1}}$ but $\Gamma>2.3$. SSRQs and NLSy1s occupy the area beyond $\Gamma>2.3$ and $L_{\gamma}>\:10^{45}\;{\rm erg\:s^{-1}}$, which is more typical for FSRQs, and are clearly separated from FRIs and FRIIs. The ratio of apparent \gray luminosity to the Eddington accretion power is different for each class of objects considered here. This ratio is  usually very small for FRIs ($<10^{-4}$), increasing for FRIIs/SSRQs ($10^{-3}-10^{-2}$), and exceeding unity for NLSY1s.
     
     \item[{\it iv)}] The \gray spectra of most of the sources studied here are described significantly better by a PL. The spectra of NGC 1275, NGC 6251, SBS 0846+513 and PMN J0948+0022 deviate from PL significantly. The spectra of NGC 6251, SBS 0846+513 and PMN J0948+0022 are better modeled with LP while the curvature in the \gray spectra of NGC 1275 can be better described using PLEC ($\beta$ free). There is a hint of possible cutoff in the spectra of 3C 380, PKS 0625-35 and 1H 0323+342, though the current data does not allow us to reject the PL model. The unusual break in the spectrum of Cen A is confirmed, which shows that below and above $2.35\pm0.08$ GeV, the photon index is $\Gamma=2.75\pm0.02$ and $2.31\pm0.1$, respectively.
    
     \item[{\it v)}] New sources having variable \gray emission have been identified. Apart rapid and strong variability of NGC 1275 and 3C 120, periods of strong flux increase and short time scale variability are found for 1H 0323+342, SBS 0846+513 and PMN J0948+0022, using the light curves generated with an adaptive binning method.
The \gray emission from 3C 380, 3C 111, Cen A core (low energy component), PKS 1502+036 and PKS 2004-447 varies within months, and there is an indication (at the threshold limit) of variability of the flux of PKS 0625-35 and M87. The \gray flux of 3C 207, 3C 275.1 and 4C +39.23B varies within half-year or one-year periods. Notable is the absence of \gray variability of Cen B and NGC 6251, which appeared to have been detected with a significance of $18.5\sigma$ and $31.6\sigma$, respectively.
    
    \item[{\it vi)}] The \gray emission from NLSy1s with their flares and peculiar multiwavelength properties makes them more  similar to blazars. The \gray emission is variable, showing flares in short time scales along with a moderate hardening of their spectra. For example, on MJD 56146.8 SBS 0846+513 was in its high state with a flux of $(4.15\pm1.04)\times10^{-7}\;{\rm photon\:cm^{-2}\:s^{-1}}$ with a significantly hard photon index of $1.73\pm0.14$. During the active state of PMN J0948+0022 from MJD 56290.96 to MJD 56294.60, the peak \gray flux was $(1.51\pm0.31)\times10^{-6}\;{\rm photon\:cm^{-2}\:s^{-1}}$ within 18.9 hrs, which corresponds to an apparent isotropic \gray luminosity of $L_{\gamma}\approx1.09\times10^{48}\;{\rm erg\;s^{-1}}$.
\end{itemize}
\section*{Acknowledgments}
This work was supported by the RA MES State Committee of Science, in the frames of the research project No 15T-1C375. Also, this work was made in part by a research grant from the Armenian National Science and Education Fund (ANSEF) based in New York, USA. We thank the anonymous referee for constructive comments that significantly improved the paper.
\bibliographystyle{aa}
\bibliography{biblio} 
\label{lastpage}
\end{document}